\documentclass[floatfix,pra,amsmath,amssymb,letterpaper,groupaddresses,superscriptaddress]{revtex4}
\usepackage{times}
\usepackage{latexsym}
\usepackage{graphicx}
\usepackage{amsmath,amssymb}
\usepackage{verbatim,times,bbm}
\usepackage{soul}
\usepackage[colorlinks,hyperindex]{hyperref}
\hypersetup{colorlinks,citecolor=blue,linkcolor=blue,urlcolor=blue}

\usepackage{color}

\newcommand{\ket}[1]{\left| #1 \right\rangle}

\begin{document}

\title{Enforcing dissipative entanglement by feedback}

\author{Morteza Rafiee\footnote{m.rafiee178@gmail.com}}
\affiliation{Faculty of Physics, Shahrood University of Technology , 3619995161 Shahrood, Iran}

\author{Alireza Nourmandipour\footnote{anourmandip@sirjantech.ac.ir}}
\affiliation{Department of Physics, Sirjan University of Technology, 7813733385 Sirjan, Iran}
\author{Stefano Mancini\footnote{stefano.mancini@unicam.it}}
\affiliation{School of Science and Technology, University of Camerino, I-62032 Camerino, Italy}
\affiliation{INFN-Sezione di Perugia, Via A. Pascoli, I-06123 Perugia, Italy}

\begin{abstract}
We study the possibility of enhancing the stationary entanglement achievable with 
two-qubit dissipating into a common environment by means of feedback.
We contrast the effect of Markovian with Bayesian feedback and 
show that, depending on the initial state, the performance of the latter
are from $16\%$ to $33\%$ superior.
\end{abstract}

\maketitle

\section{Introduction}

Over recent years it became evident that quantum entanglement is a key resource for many applications of quantum information theory \cite{NCh00}. Therefore many attempts, both theoretically \cite{BR19} and experimentally \cite{ZD16}, have been devoted to find schemes to produce entangled states. In this regard the life time of entangled states was a central issue,
because dissipative effects occurring in real world tend to wash out entanglement. 
However, this fact has been proven not to be necessarily true when constituents of a system, like qubits, are coupled to a common environment. Recent studies put forward the idea that dissipative dynamics can induce stationary entanglement without direct interaction of the subsystems \cite{MM11,NTR16,KM11,MM13,HKD09,NT15}. Nevertheless the amount of such entanglement is usually quite low, therefore it seems natural to seek for strategies to control and even enhance it. 
 
The possibility to use information gathered from measurements to alter the dynamics of a quantum system, i.e. the realization of quantum feedback sounds palatable to this end. Actually, quantum feedback has been already proposed  
for controlling entanglement degradation in interacting systems (see e.g. \cite{WWM05,MW05,RNM16,RNM17}). Interestingly, these strategies can be implemented in several experiments \cite{ZHLWJN17}. 
They are however based on the direct (without processing) usage of the information lost into environment and recovered as classical current. Thus referred to as Markovian quantum feedback \cite{WM93}.
Although such a kind of feedback has found many successful applications (such as noice reduction and quantum error correction \cite{CLG08}, state reduction and stabilization \cite{WW01}, quantum state discrimination \cite{HBD09}, quantum parameter estimation \cite{SGDM04})
it is quite far from being optimal. 

A clever feedback control is based on a suitable (filtering) process of the measurement record, ideally on an estimate of the system state \cite{DJ99}. Hence the name Bayesian feedback \cite{WMW02}. 
Since the filtering process can be regarded as a learning process, the Bayesian quantum feedback control can also be assimilated to quantum machine learning \cite{SSP15}. 
In the seminal work \cite{WMW02}, it has been shown that such a feedback is always superior to the Markovian one for stabilizing qubit states. The nonlinear nature of qubits dynamics in presence of an environment gives rise to many difficulties in dealing with the Bayesian feedback control and suppressed developments even for two-qubit systems. 

Here, we consider a Bayesian feedback mechanism to control and enhance the stationary entanglement between two qubits dissipating into a common environment, thus involving 
nonlocal measurements. We assume to solve in real time the nonlinear stochastic equations to obtain the conditioned state and based on that we realize \emph{local actuations} as classical drivings.
We show that such a feedback scheme works well even if the qubits are non directly interacting.  
Above all, we show a great advantage in terms of performance in using Bayesian feedback with respect to Markovian one. This however depends on the initial state and is larger when the latter lies in the symmetric subspace. 

The rest of the paper is organized as follow. We start by introducing the model and its uncontrolled dynamics in Sec.\ref{sec:model}.
Then we deal with the Markovian feedback control in Sec. \ref{sec:Markovian}. Section \ref{sec:Bayesian} presents the Bayesian feedback control. Finally, Sec. \ref{sec:conclu} is for conclusion.
Throughout the paper we will use $\iota$ to denote the imaginary unit.

%%%%%%%%%%%%%%%%%%%%%%%%%%%%%%%%%%%%%%%%%%%%%%%%%%

\section{The model}\label{sec:model}

Let us consider a system composed by two qubits 
with associated Pauli operators $\sigma_j^{x,y,z}$
and lowering, rising operators $\sigma_j,\sigma_j^\dag$, $j=1,2$.
Suppose that the two qubits dissipate into a common environment. 
According to the prescriptions of \cite{BP02},
their (uncontrolled) dynamics will be described by the following master equation
\begin{eqnarray}\label{master}
{\dot{\rho}} =-\imath \,  \left[ H,\rho\right]+{\mathcal D}[\Sigma]\rho,
\end{eqnarray}
in which $\rho$ is the system density operator, $H$ the Hamiltonian, and ${\mathcal D}$ is the dissipative superoperator whose action depends on the system operator $\Sigma=\sigma_1+\sigma_2$ that determines the coupling to the environment
\begin{eqnarray}
{\mathcal D}[\Sigma]\rho\equiv\Sigma \rho \Sigma^\dagger
-\frac{1}{2}\left(\Sigma^\dagger \Sigma \rho+\rho \Sigma^\dagger \Sigma\right).
\end{eqnarray}
Actually this is a special case of a more general model in which the two qubits can dissipate into both local and global environments at nonzero temperatures \cite{NTM16}. 

There are several approaches to unravel the master equation (\ref{master}). For instance, it can be assumed that the environment makes (continuous) measurements on the system. However, the results of each measurement are quickly forgotten due to rapid thermalization \cite{ADL02}. Therefore, the non-selective evolution of the system is the average over all possible measurements. It is then straightforward to show that the continuous measurements performed by the environment with the following (for example) Kraus operators
\begin{subequations}
\begin{eqnarray}
\Omega_0(dt)&\equiv&1-\left(\iota H+\frac{1}{2}\Sigma^\dag \Sigma\right)dt,\\
\Omega_1(dt)&\equiv&\Sigma\sqrt{dt}
\end{eqnarray}
\end{subequations}
is equivalent to the master equation (\ref{master}). 
The above consideration suggests that the selective measurement gives rise to a trajectory 
in the space of the state matrices $\rho$ (rather than in the Hilbert space) \cite{WMW02}. 
Therefore, the unravelled master equation (\ref{master}) can be thought as an average over many trajectories, each one given by the following stochastic master equation (SME) \cite{WD05}
\begin{equation}
\label{SME2}
d\rho^{(I)}=\left\{-\imath \left[ H,\rho^{(I)}\right]+{\mathcal D}[\Sigma]\rho^{(I)} \right\} dt
+{\cal H}[\Sigma]\rho^{(I)} dW(t),
\end{equation}
where
\begin{equation}
{\cal H}[\Sigma]\rho^{(I)}\equiv\left(\Sigma\rho^{(I)}+\rho^{(I)}\Sigma^\dag\right) 
-{\rm Tr}\left(\Sigma\rho^{(I)}+\rho^{(I)}\Sigma^{\dagger}\right) \, \rho^{(I)},
\end{equation}
$dW$ is the infinitesimal Wiener increment \cite{CWG1985}
defined by $\mathbb{E}(dW)=0$, $dW^2=dt$,
and $\rho^{(I)}$ is the state conditioned to the measurement current
\begin{equation}\label{homodyne}
I(t) dt={\rm Tr}\left(\Sigma\rho^{(I)}+\rho^{(I)}\Sigma^{\dagger}\right)dt +dW.
\end{equation}
Although taking the ensemble average $\mathbb{E}$ (over all possible measurement realizations) of \eqref{SME2} we will recover \eqref{master}, this point of view is of great importance for implementing control actions. 

The Hamiltonian $H$ will be considered as as not having any interaction (non-local) term,
while for local terms we assume  
the two qubits driven by a resonant classical field in the $x$-direction. Therefore,  
\begin{equation}
H=\omega\Sigma_x=\omega(\Sigma+\Sigma^\dag)
=\omega(\sigma_1^x+\sigma_2^x), 
\end{equation}
where $\omega$ is the (real) amplitude of the driving field, so that the deterministic master equation \eqref{master} becomes
\begin{eqnarray}\label{masterfin}
{\dot{\rho}} =-\imath \, \omega  \left[ \Sigma_x,\rho\right]+{\mathcal D}[\Sigma]\rho.
\end{eqnarray}
We are interested in the stationary solutions of Eq. (\ref{masterfin}). It should be noted that the steady state is not unique and will depend on the initial state. This is due to the fact that there exist non-trivial operators (i.e., not multiple of the identity) commuting with the Lindblad operators $\Sigma$ \cite{Spohn}.
Actually, introducing the computational basis $\{|11\rangle,|10\rangle,|01\rangle,|00\rangle\}$
where $|0\rangle$ (resp. $|1\rangle$) donotes the ground (resp. excited) state, any operator proportional to
\begin{eqnarray}\label{operator}
\left(
\begin{array}{cccc}
 2 &  0 & 0 & 0 \\
  0 &  1 & 1 & 0 \\
  0 &  1 & 1 & 0 \\
  0 &  0 & 0 & 2
\end{array}\right)
\end{eqnarray}
commutes with the Lindblad operator $\Sigma$.

In order to solve the master equation \eqref{masterfin}, we first write the density operator 
in the matrix form, with the following unknown time dependent entries, in the computational basis
\begin{equation}
\rho=\left(
\begin{array}{cccc}
{\cal A} & {\cal B}_R+\iota {\cal B}_I & {\cal C}_R+\iota {\cal C}_I & {\cal D}_R+\iota {\cal D}_I \\ 
{\cal B}_R-\iota {\cal B}_I & {\cal E} & {\cal F}_R+\iota {\cal F}_I & {\cal G}_R+\iota {\cal G}_I \\ 
{\cal C}_R-\iota {\cal C}_I & {\cal F}_R-\iota {\cal F}_I & {\cal H} & {\cal I}_R+\iota {\cal I}_I \\ 
{\cal D}_R-\iota {\cal D}_I & {\cal G}_R-\iota {\cal G}_I & {\cal I}_R-\iota {\cal I}_I & 1-{\cal A}- {\cal E}-{\cal H}
\end{array}\right).
\label{eq:rhomatrix}
\end{equation}
and then vectorize it as 
\begin{equation}
\label{Eq:V}
{\bf v}\equiv  
\left( 
\begin{array}{ccccccccccccccc} 
{\cal A} & {\cal B}_R & {\cal B}_I & {\cal C}_R & {\cal C}_I & {\cal D}_R & {\cal D}_I & {\cal E} 
& {\cal F}_R & {\cal F}_I & {\cal G}_R & {\cal G}_I & {\cal H} & {\cal I}_R & {\cal I}_I 
\end{array}
\right)^\top,
\end{equation}
so that Eq. \eqref{masterfin} becomes
\begin{equation}\label{eqmatrixnew1}
\dot{\bf v}={\bf M}{\bf v}-{\bf w},
\end{equation}
with matrix ${\bf M}$ and vector ${\bf w}$ of constant coefficients (see Appendix \ref{appA}).
Using the Laplace transform we get
\begin{equation}\label{eqmatrixnew1Lap}
s \tilde{\boldsymbol{v}}(s)-{\bf v}(0)={\bf M}\tilde{\boldsymbol{v}}(s)-\tilde{\boldsymbol{w}}(s),
\end{equation}
where $\tilde{\boldsymbol{v}}(s)$  and $\tilde{\boldsymbol{w}}(s)$ are the vectors whose entries are the Laplace transformed entries of ${\bf v}(t)$ and ${\bf w}$, respectively. Then, the set of algebraic equations coming from \eqref{eqmatrixnew1Lap} can be analytically solved. Since we are only interested in the stationary solution of \eqref{eqmatrixnew1} (in turn \eqref{masterfin}), we use the Final Value Theorem \cite{Graf2004} to obtain the following analytical expression for the stationary state:
\begin{equation}\label{eq:rhomatrixSta}
\rho(\infty)=\frac{R}{2} \rho_s + \frac{2-R}{2}\rho_a,
\end{equation}
where
\begin{equation}
\rho_s\equiv\frac{1}{12\omega^4+4\omega^2+1}\left(
\begin{array}{cccc}
 4 \omega ^4 & -2 i \omega ^3 & -2 i \omega ^3 & -2\omega ^2 \\
 2 i \omega ^3 & 2 \omega ^4+\omega ^2 & 2 \omega ^4+\omega ^2 & 
 -i \omega\left(2 \omega ^2+1\right) \\
 2 i \omega ^3 & 2 \omega ^4+\omega ^2 & 2 \omega ^4+\omega ^2 & 
 -i \omega  \left(2 \omega ^2+1\right) \\
 -2 \omega ^2 & i \omega  \left(2 \omega ^2+1\right) & 
 i \omega\left(2 \omega ^2+1\right) & 4 \omega ^4+2 \omega ^2+1
\end{array}
\right),
\label{eq:rhomatrixStaFixed}
\end{equation}
\begin{equation}
\rho_a\equiv\frac{1}{2}\left(
\begin{array}{cccc}
 0 & 0 & 0 & 0 \\
 0 & 1 & -1 & 0 \\
 0 & -1 & 1 & 0 \\
 0 & 0 & 0 & 0 \\
\end{array}
\right),
\label{eq:V}
\end{equation} 
with 
\begin{equation}
R\equiv 2+2{\cal F}_R(0)-{\cal H}(0)-{\cal E}(0).
\end{equation}
It is clear from \eqref{eq:rhomatrixSta} that there are two contributions to the stationary state.
The first one $\rho_s$ is related to the symmetric subspace spanned by $\{ |11 \rangle, 
\frac{1}{\sqrt{2}}(|10 \rangle+|01 \rangle),|00\rangle\}$,
while the second one $\rho_a$ is related to the antisymmetric subspace spanned by 
$\{\frac{1}{\sqrt{2}}(|10 \rangle-|01 \rangle)\}$.
This can be understood by noting that the Hamiltonian $\omega\Sigma_x$ as well as dissipative term ${\cal D}[\Sigma]$ do not mix symmetric and antisymmetric subspaces. As consequence the dynamical map  \eqref{masterfin} has a fixed point $\rho_s$ when acting on the symmetric subspace ($R=2$) and a fixed point $\rho_a$ when acting on the antisymmetric subspace ($R=0$). Therefore, when it acts on an initial state that has components in both symmetric and antisymmetric subspaces, we get the convex combination \eqref{eq:rhomatrixSta} of $\rho_s$ and $\rho_a$.

\medskip

The stationary entanglement is quantified by means of concurrence \cite{WKW98}
\begin{equation}
C(\rho)\equiv \max\{0,\lambda_1-\lambda_2-\lambda_3-\lambda_4\},
\end{equation}
where $\lambda_i$ are, in decreasing order, the non-negative square roots of the moduli of the eigenvalues of 
\begin{equation}
\rho \left((\sigma_1-\sigma_1^\dag)\otimes (\sigma_2-\sigma_2^\dag)\right)
\rho^* \left((\sigma_1-\sigma_1^\dag)\otimes (\sigma_2-\sigma_2^\dag)\right).
\end{equation}
Explicitly the concurrence will be function of the parameters characterizing the state, i.e. 
the (real) driving field amplitude $\omega$ and the feedback strength (see following Sections)
$\lambda$.  We henceforth denote
 \begin{equation}
\hat{C}(\omega)\equiv\max_{\lambda}C(\omega,\lambda),
\qquad
\hat{\lambda}\equiv\arg\max_{\lambda}C(\omega,\lambda).
\end{equation}
Fig.\ref{Fig1}, where it is reported $C(\omega,0)$ vs $\omega$, illustrates that a tiny amount of entanglement can be achieved for nonzero driving (maximum concurrence $\approx 0.11$ at  $\omega\approx 0.4$).
This result holds for all initial states in the symmetric subspace.

%%%%%%%%%%%%%%%%%%%%%%%%%%%%%%%%%%%%%%%%%%%%%%%%%%

\section{Markovian feedback}\label{sec:Markovian}

Markovian feedback amounts to directly use the measurement current $I(t)$ in \eqref{homodyne}
to drive the system \cite{HMW}. This can be described by means of a feedback 
Hamiltonian of the form 
\begin{equation}\label{HMar}
H_{\text{fb}}=\lambda I(t) F(t),
\end{equation}
where $F(t)$ is a Hermitian operator and $\lambda$ is a constant characterizing the feedback strength.

As a consequence of \eqref{HMar},  the SME \eqref{SME2} becomes 
\begin{align}\label{SMEMar}
d\rho^{(I)}(t)&=\left\{-\imath\,  \left[ H,\rho^{(I)}\right]+{\mathcal D}[\Sigma]\rho^{(I)} 
-\imath \left[ \lambda F, \Sigma\rho^{(I)}+\rho^{(I)}\Sigma^\dag\right]
+{\cal D}[\lambda F] \rho^{(I)}
\right\} dt \notag\\
&+\left\{{\cal H}[\Sigma-\imath \lambda F]\rho^{(I)} \right\} dW(t),
\end{align}
which is a true Ito equation with $dW(t)$ independent of $\rho^{(I)}(t)$ \cite{HMW}.
Thus one can eventually take also the ensemble average and obtain 
\begin{align}\label{MEMar}
\dot{\rho}=-\imath\,  \left[ H+\frac{\lambda}{2}\left(  \Sigma^\dag F+F\Sigma\right) ,\rho\right]
+{\mathcal D}[\Sigma-\imath \lambda F]\rho.
\end{align}
It is evident that the effect of the feedback is to replace $\Sigma$ by $\Sigma-\imath\lambda F$ and the addition of the extra term $\frac{\lambda}{2}\left(  \Sigma^\dag F+F\Sigma\right)$ to the Hamiltonian. The above master equation turns out to still be of Lindblad and Markovian form \cite{BP02}. This is due to the fact that the information gathered through $I(t)$ is instantaneously fed back to the system, i.e. with no time delays.

In the previous section, we have considered the driving Hamiltonian to be polarized along the $x$-direction, i.e. $\omega\Sigma_x$, which makes each qubit to undergo local oscillations between its energy eiegenstates. These oscillations should have a fixed phase relationship with the driving provided by the feedback Hamiltonian in order for the latter to be effective.
Therefore, we assume $H_{\text{fb}}$ as well polarized  along the $x$-direction. In particular, with the choice $H_{\text{fb}}=\lambda I(t) \Sigma_x$,  Eq. (\ref{MEMar}) becomes
\begin{align}\label{SMEMarfeed}
\dot{\rho}=-\imath\, \omega  \left[ \Sigma_x,\rho\right]
-\imath \frac{\lambda}{2}  \left[ \Sigma^\dag \Sigma_x+\Sigma_x\Sigma,\rho\right]
+{\mathcal D}[\Sigma-\imath \lambda \Sigma_x]\rho.
\end{align}

Writing again $\rho$ as (\ref{eq:rhomatrix}), Eq. (\ref{SMEMarfeed}) can be put in the same form of (\ref{eqmatrixnew1}), where now {\bf M} and {\bf w} are given in Appendix \ref{appB}. 
After taking the same steps as in Sec. \ref{sec:model}, it is possible to arrive to the same formal solution \eqref{eq:rhomatrixSta}, where however now
\begin{equation}\label{eq:rhomatrixStaFeed}
\rho_s=\frac{1}{T}\left(
\begin{array}{cccc}
 \Upsilon_1 & \Upsilon_2+\imath\Upsilon_3 & \Upsilon_2+\imath\Upsilon_3 & \Upsilon_4+\imath\Upsilon_5 \\
 \Upsilon_2-\imath\Upsilon_3 & \Upsilon_6 & \Upsilon_6 & \imath\Upsilon_7 \\
 \Upsilon_2-\imath\Upsilon_3 & \Upsilon_6 & \Upsilon_6 & \imath\Upsilon_7 \\
 \Upsilon_4-\imath\Upsilon_5 & -\imath\Upsilon_7 & -\imath\Upsilon_7 & 1-\Upsilon_1-2\Upsilon_6 \\
\end{array}
\right),
\end{equation}
with parameters $T$ and $\Upsilon_i$ given in Appendix \ref{appB}. 
We point out that taking the limit $\lambda\to 0$ in \eqref{eq:rhomatrixStaFeed} we readily obtain Eq. (\ref{eq:rhomatrixStaFixed}).

\begin{figure}[!ht]
\centering
	\includegraphics[width=9cm,height=7cm,angle=0]{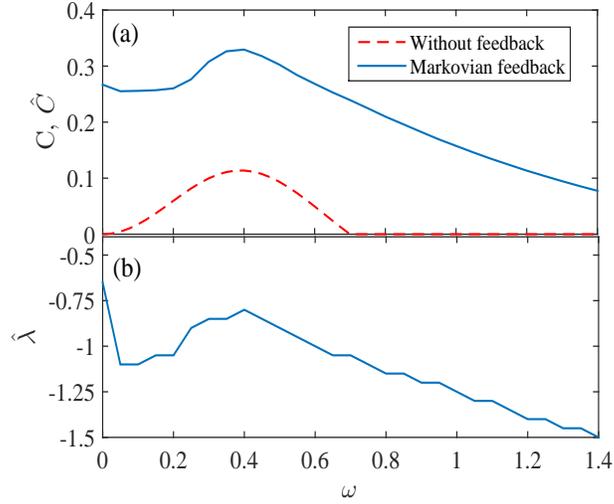}
	\caption{ (a) Stationary entanglement $C(\omega,\lambda=0)$ (red-dashed line) and maximum stationary entanglement under Markovian feedback $\hat{C}(\omega)$ (blu-solid line)
	vs driving amplitude $\omega$.
	 (b) Value of the optimal feedback amplitude $\hat{\lambda}$ vs $\omega$.
	 Plots can be referred to any initial state in the symmetric subspace.}
	\label{Fig1}
\end{figure}

In Fig. \ref{Fig1} (a), we illustrate the stationary entanglement for the case without feedback, i.e., $C(\omega,0)$ (dashed line) together with the maximum value of entanglement achievable with Markovian feedback  $\hat{C}(\omega)$ (solid line) vs $\omega$. In Fig. \ref{Fig1} (b),  it is reported the optimal values of feedback amplitude $\hat{\lambda}$ vs $\omega$.
These plots are valid for any initial state in the symmetric subspace.
The maximum amount of stationary entanglement with feedback, 0.31, is achieved with a driving amplitude $\omega = \approx 0.4$ and feedback amplitude $\lambda=-0.8$.  This is in agreement with the results of \cite{WWM05}, where Markovian fedback was investigated in symmetric subspace.
Notice that the plots can be mirrored to negative values of $\omega$.
It is worth remarking the enhancement of stationary entanglement (from 0.11 to 0.31) already obtainable with Markovian feedback. 

\begin{figure}[!ht]
	\centering
	\includegraphics[width=9cm,height=7cm,angle=0]{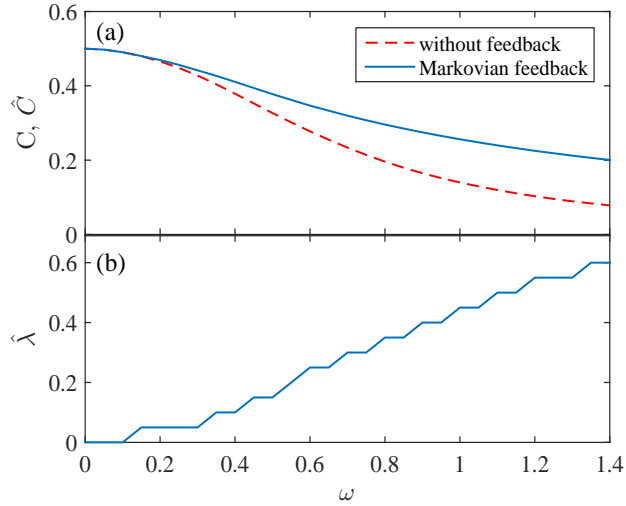}
	\caption{ (a) Stationary entanglement $C(\omega,\lambda=0)$ (red-dashed line) and maximum stationary entanglement under Markovian feedback $\hat{C}(\omega)$ (blu-solid line)
		vs driving amplitude $\omega$.
		(b) Value of the optimal feedback amplitude $\hat{\lambda}$ vs $\omega$.
		In both plots $|10 \rangle$ is the initial state.}
	\label{Fig2}
\end{figure}

 Fig. \ref{Fig2} illustrates the results when the the initial state has a nonzero component also in the antisymmetric subspace, specifically it is $|10\rangle$. 
Actually, Fig. \ref{Fig2} (a) shows that Markovian feedback is not effective in this case, since the value of 
$\hat{C}(\omega)$ never exceeds that of $C(\omega,\lambda=0)$.

%%%%%%%%%%%%%%%%%%%%%%%%%%%%%%%%%%%%%%%%%%%%%%%%%%

\section{Bayesian feedback}\label{sec:Bayesian}

We now consider controlling the system dynamics using a
Hamiltonian that depends not directly on the current, but
rather on the observer's state of knowledge of the system $\rho^{(I)}$.
By definition there is nothing better with which to control the
system. Taking into account that the actuation 
should result in a driving proportional to $\Sigma_x$, we have in general
\begin{equation}\label{HBay}
H_{fb}=f\left(\rho^{(I)}\right) \Sigma^x,
\end{equation}
where $f$ is an arbitrary function of $\rho^{(I)}$.
Then, using \eqref{HBay} in the SME \eqref{SME2} we get
\begin{equation}\label{SMEBayes}
d\rho^{(I)}=\left\{-\imath\, \omega  \left[ \Sigma^x,\rho^{(I)}\right]+{\mathcal D}[\Sigma]\rho^{(I)} 
-\imath f \left[ \Sigma^x,\rho^{(I)}\right] 
\right\} dt
+{\cal H}[\Sigma]\rho^{(I)} dW(t).
\end{equation}
Assuming perfect knowledge of the system dynamics,
this is a nonlinear stochastic Markovian equation.
However, it is not possible to average over the stochasticity
to obtain a master equation. This reveals the underlying non-Markovicity of Bayesian feedback.

For what concern the function $f\left(\rho^{(I)}\right)$ we intuitively opt
for the following one
\begin{equation}\label{fBay}
f(\rho^{(I)}(t))\equiv\lambda \, {\rm sgn} \left[ C\left(\rho^{(I)}(t)\right)-C\left(\rho^{(I)}(t- \Delta t)\right)\right],
\end{equation}
where $\Delta t$ is an infinitesimal time step. 
The effect would be the following:
if the concurrence is increased in the latest time step, then one continues to drive the system with a force having the same sign as before, otherwise if the concurrence is decreased in the latest time step, one changes the sign of the driving force.

 Equation \eqref{SMEBayes} has been solved numerically by using Euler-Maruyama method, which is specific for numerical solution of a stochastic differential equations \cite{Higham}. To apply it we chose a very small time step $\Delta t=10^{-6}$ and we obtain the solution $\rho^I(t)$ over $5*10^3$ realizations. Then the average overall these trajectories leads to a stable convergent  solution for unconditional density matrix. 

It is to note that with this (Bayesian) feedback, the stationary state depends on the initial one no matter whether the latter is or not in the symmetric subspace.
 
 \begin{figure}[!ht]
 \centering
  	\includegraphics[width=9cm,height=7cm,angle=0]{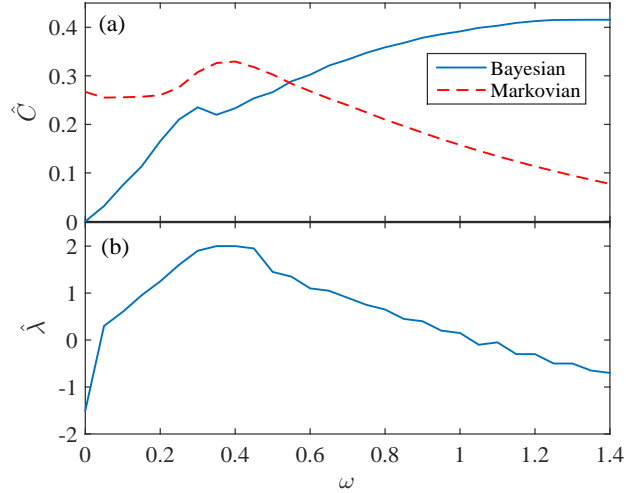},
  	\caption{ (a) Maximum stationary entanglement $\hat{C}(\omega)$ obtained from Bayesian feedback (blu-solid line) and Markovian feedback (red-dashed line) vs $\omega$. (b) Value of optimal feedback amplitude $\hat{\lambda}$ vs $\omega$ for Bayesian feedback. The initial state is $|00 \rangle$.}
  	\label{Fig3}
  \end{figure}

In Fig. \ref{Fig3} we considered the initial state  $|00\rangle$ and we contrasted the maximum stationary entanglement  $\hat{C}(\omega)$ resulting from Bayesian and Markovian feedback.
It is evident that the Bayesian feedback starts to become effective as long as $\omega$ increases from zero, while Markovian feedback is effective also at $\omega=0$. This is due to the fact that for 
$\omega=0$ the state  $|00\rangle$ remains unaltered under uncontrolled (even conditioned) dynamics. Then the driving of Markovian feedback, being proportional to $\Sigma_x$ is effective, while the driving of Bayesian feedback being also proportional to the concurrence, is not effective.  
For large values of $\omega$ the entanglement by Bayesian feedback tends to the asymptotic value of $0.41$, which is quite larger than the maximum $0.31$ achievable by Markovian feedback.
  
 \begin{figure}[!ht]
  \centering
 	\includegraphics[width=9cm,height=7cm,angle=0]{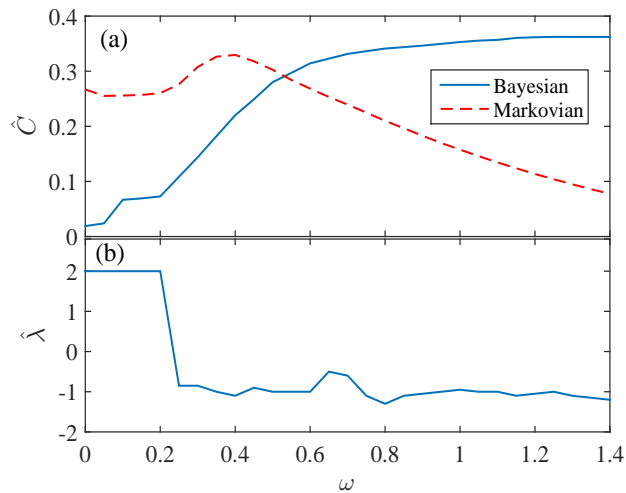},
 	\caption {(a) Maximum stationary entanglement $\hat{C}(\omega)$ obtained from Bayesian feedback (blu-solid line) and Markovian feedback (red-dashed line) vs $\omega$. (b) Value of optimal feedback amplitude $\hat{\lambda}$ vs $\omega$ for Bayesian feedback. The initial state is 
	$|11\rangle$.}
 	\label{Fig4}
 \end{figure}
 
 In Fig. \ref{Fig4} we considered the initial state  $|11\rangle$ and we contrasted the maximum stationary entanglement  $\hat{C}(\omega)$ resulting from Bayesian and Markovian feedback.
Clearly the entanglement from Markovian feedback is the same as in Fig. \ref{Fig2} since both initial states  $\ket{00}$ and $\ket{11}$ live in the symmetric subspace.
Also the behavior of entanglement from Bayesian feedback is similar to that of Fig. \ref{Fig2}, however in this case there is a nonzero amount at $\omega=0$.  This is due to the fact that the initial state have excitations, hence the conditioned state results entangled.
 For large values of $\omega$ the entanglement by Bayesian feedback tends to the asymptotic value of $0.36$, which is still larger than the maximum $0.31$ achievable by Markovian feedback.
  
 \begin{figure}[!h]
  \centering
 	\includegraphics[width=9cm,height=7cm,angle=0]{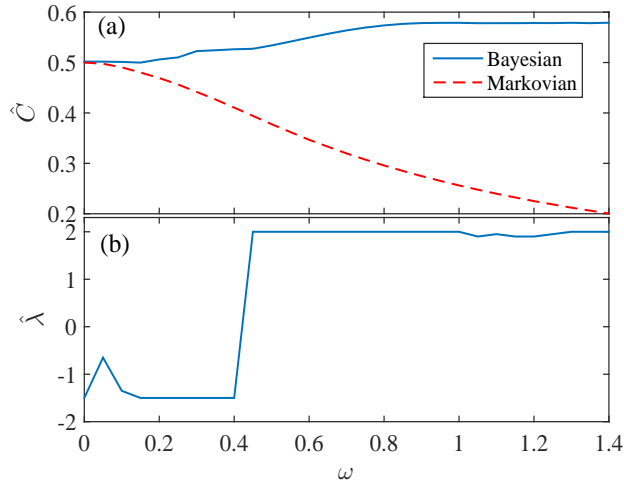},
 	\caption{ (a) Maximum stationary entanglement $\hat{C}(\omega)$ obtained from Bayesian feedback (blu-solid line) and Markovian feedback (red-dashed line) vs $\omega$. (b) Value of optimal feedback amplitude $\hat{\lambda}$ vs $\omega$ for Bayesian feedback. The initial state is 
	$|10 \rangle$.}
 	\label{Fig5}
 \end{figure}
 
  Finally, Fig. \ref{Fig5} deals with an initial state  $\ket{10}$ not belonging to the symmetric subspace (and of course not belonging to the antisymmetric subspace too).
  Also in this case we contrasted the maximum stationary entanglement  $\hat{C}(\omega)$ resulting from Bayesian and Markovian feedback.
They both start from the same value $0.5$ at $\omega=0$, but then while the Markovian feedback is monotonically decreasing, the Bayesian feedback is monotonically increasing till the asymptotic value $\approx 0.58$. Again this latter is larger than the maximum 0.5 of the Markovian feedback.

At the end we remark that in all cases the optimal feedback action is achieved with finite value of the feedback strength $\hat{\lambda}$.

%%%%%%%%%%%%%%%%%%%%%%%%%%%%%%%%%%%%%%%%%%%%%%%%%%

\section{Conclusion}\label{sec:conclu}

To sum up, we have investigated feedback mechanisms in two qubits dissipating into a common environment with the aim of controlling and enhancing their stationary entanglement. We started from the uncontrolled system and derived the analytical expression for the stationary state. In this case the amount of the optimal stationary entanglement is about $0.11$ for initial states living in the symmetric subspace.  Since dissipation occurs in common environment the corresponding measured system operator is $\Sigma_x$ and hence the best local control turns out to be a driving proportional to such an operator.  Then we illustrated that the amount of entanglement can be enhanced up to $0.31$ by considering Markovian feedback control.  This is true for all initial states living in the symmetric subspace. However for states not fully contained in the symmetric subspace the Markovian feedback loses its effectiveness.

To obtain more general and better results we considered a filtering process on the measurement record. With this we realized a state estimation to determine the change in the amount of entanglement at each time step and then adjusted consequently the driving.   Such kind of Bayesian feedback control, unlike the Markovian one,  always depends on the initial state (even for the states living in the symmetric subspace) and we found its performance to be from 16\% to 33\% superior.
Interestingly the Bayesian feedback is effective also with initial states not fully contained in the symmetric subspace. Moreover, the optimal feedback always occurs with finite strength.
The assumption of solving the non-linear stochastic equation \eqref{SMEBayes} in real time can be reasonable with the use of powerful computer for which integration time can be smaller than the typical dynamical timescales of the system. Of course there can be other imperfections reducing the performance, like inefficient detection \cite{WM2010}, but this affects both kinds of feedback.

The realization of the proposed control scheme seems to be within reach by present day technologies. For instance, using a fast, ultralow-noise parametric amplifier allows us to observe the quantum jumps between qubit states in real-time \cite{VSS11,VMS13}. More precisely, our quantum system could be considered as two anharmonic oscillators realized by capacitively shunted Josephson junction coupled to a three-dimensional microwave cavity. Then the two lowest energy levels of such oscillators can form our transmon qubits.
The cavity can be probed with photons in order to realize weak measurement of the qubits. The photons are then directed to a high-bandwidth, quantum-noise-limited amplifier, to have a real-time monitoring of the cavity state. This, in turn, will be used to modulate the amplitude of the Rabi driving signal \cite{VMS12}.

Finally, although by definition of Bayesian feedbcak there is nothing better with which to control the system, we do not have a rigorous proof of the optimality for the specific choices of 
\eqref{HBay} and \eqref{fBay}, 
hence a future study could address this issue exploiting some suitable criteria \cite{JSh08}.

%%%%%%%%%%%%%%%%%%%%%%%%%%%%%%%%%%%%%%%%%%%%%%%%%%

\appendix

\section{Dynamics without feedback}\label{appA}

It is
\begin{equation}
\dot{\bf v}={\bf M}{\bf v}-{\bf w},
\end{equation}
where
\begin{equation}
\label{Wnofeedback}
{\bf w}:=\left(
\begin{array}{ccccccccccccccc}
0 & 0 & 0 & 0 & 0 & 0 & 0 & 0 & 0 & 0 & 0 & \omega & 0 & 0 & \omega
\end{array}\right)^\top,
\end{equation}
and
\begin{eqnarray}
{\bf M}=
\left(
\begin{array}{ccccccccccccccc}
 -2 & 0 & -2 \omega  & 0 & -2 \omega  & 0 & 0 & 0 & 0 & 0 & 0 & 0 & 0 & 0 & 0 \\
 0 & -\frac{3}{2} & 0 & -\frac{1}{2} & 0 & 0 & -\omega  & 0 & 0 & -\omega  & 0 & 0 & 0 & 0 &
   0 \\
 \omega  & 0 & -\frac{3}{2} & 0 & -\frac{1}{2} & \omega  & 0 & -\omega  & -\omega  & 0 & 0 &
   0 & 0 & 0 & 0 \\
 0 & -\frac{1}{2} & 0 & -\frac{3}{2} & 0 & 0 & -\omega  & 0 & 0 & \omega  & 0 & 0 & 0 & 0 &
   0 \\
 \omega  & 0 & -\frac{1}{2} & 0 & -\frac{3}{2} & \omega  & 0 & 0 & -\omega  & 0 & 0 & 0 &
   -\omega  & 0 & 0 \\
 0 & 0 & -\omega  & 0 & -\omega  & -1 & 0 & 0 & 0 & 0 & 0 & \omega  & 0 & 0 & \omega  \\
 0 & \omega  & 0 & \omega  & 0 & 0 & -1 & 0 & 0 & 0 & -\omega  & 0 & 0 & -\omega  & 0 \\
 1 & 0 & 2 \omega  & 0 & 0 & 0 & 0 & -1 & -1 & 0 & 0 & -2 \omega  & 0 & 0 & 0 \\
 1 & 0 & \omega  & 0 & \omega  & 0 & 0 & -\frac{1}{2} & -1 & 0 & 0 & -\omega  & -\frac{1}{2}
   & 0 & -\omega  \\
 0 & \omega  & 0 & -\omega  & 0 & 0 & 0 & 0 & 0 & -1 & \omega  & 0 & 0 & -\omega  & 0 \\
 0 & 1 & 0 & 1 & 0 & 0 & \omega  & 0 & 0 & -\omega  & -\frac{1}{2} & 0 & 0 & -\frac{1}{2} &
   0 \\
 \omega  & 0 & 1 & 0 & 1 & -\omega  & 0 & 2 \omega  & \omega  & 0 & 0 & -\frac{1}{2} &
   \omega  & 0 & -\frac{1}{2} \\
 1 & 0 & 0 & 0 & 2 \omega  & 0 & 0 & 0 & -1 & 0 & 0 & 0 & -1 & 0 & -2 \omega  \\
 0 & 1 & 0 & 1 & 0 & 0 & \omega  & 0 & 0 & \omega  & -\frac{1}{2} & 0 & 0 & -\frac{1}{2} & 0
   \\
 \omega  & 0 & 1 & 0 & 1 & -\omega  & 0 & \omega  & \omega  & 0 & 0 & -\frac{1}{2} & 2
   \omega  & 0 & -\frac{1}{2} \\
\end{array}
\right).
\label{M0}
\end{eqnarray}

%%%%%%%%%%%%%%%%%%%%%%%%%%%%%%%%%%%%%%%%%%%%%%%%%

\section{Solution of the dynamics with Markovian feedback }\label{appB}

It is again
\begin{equation}
\dot{\bf v}={\bf M}{\bf v}-{\bf w},
\end{equation}
where however
\begin{equation}
\label{Wfeedback}
{\bf w}:=\left(
\begin{array}{ccccccccccccccc}
0 & 0 & 0 & 0 & 0 & \lambda^2 & 0 & -\lambda^2 & -\lambda^2 & 0 & 0 & \omega & -\lambda^2 & 0 & \omega
\end{array}\right)^\top,
\end{equation}
and
\begin{equation}
{\bf M}\equiv \left({\bf M}_1 \, {\bf M}_2 \, {\bf M}_3 \right)
\end{equation}
with
\begin{eqnarray}
{\bf M}_1\equiv\left(
\begin{array}{ccccc}
 2 A(2) & 0 & -2 \omega  & 0 & -2 \omega   \\
 0 & A(3) & 0 & -\frac{1}{2} & -2 \lambda   \\
 \omega  & -2 \lambda  & C(0) & 0 & B(1) \\
 0 & -\frac{1}{2} & -2 \lambda  & A(3) & 0  \\
 \omega  & 0 & B(1) & -2 \lambda  & C(3) \\
 0 & 0 & -\omega  & 0 & -\omega  \\
 2 \lambda  & \omega  & 0 & \omega  & 0 \\
 1 & 0 & 2 \omega  & 0 & 0  \\
 1 & 0 & \omega  & 0 & \omega  \\
 0 & \omega  & 0 & -\omega  & 0  \\
 0 & 1 & 2 \lambda  & -A(2) & 0  \\
 \omega  & 2 \lambda  & -B(2) & 0 & -A(2) \\
 1 & 0 & 0 & 0 & 2 \omega  \\
 0 & -A(2) & 0 & 1 & 2 \lambda \\
 \omega  & 0 & -A(2) & 2 \lambda  & -B(2) 
\end{array}
\right),
\label{M1}
\end{eqnarray}

\begin{eqnarray}
{\bf M}_2\equiv\left(
\begin{array}{ccccc}
B(0) & 0 & -A(0) & -B(0) & 0  \\
 0 & -\omega  & 0 & 0 & -\omega \\
 \omega  & 0 & -\omega  & -\omega  & 0  \\
 0 & -\omega  & 0 & 0 & \omega \\
 \omega  & 0 & 0 & -\omega  & 0 \\
 B(2) & 2 \lambda  & -B(0) & -B(0) & 0 \\
 -2 \lambda  & B(2) & -\lambda  & -2 \lambda  & 0 \\
 -B(0) & -2 \lambda  & C(2) & B(2) & -2 \lambda  \\
-B(0) & -2 \lambda  & B(1) & B(2) & 0  \\
 0 & 0 & \lambda  & 0 & B(2) \\
0 & \omega  & 0 & 0 & -\omega \\
 -\omega  & 0 & 2 \omega  & \omega  & 0  \\
 -B(0) & -2 \lambda  & A(0) & B(2) & 2 \lambda  \\
 0 & \omega  & 0 & 0 & \omega  \\
 -\omega  & 0 & \omega  & \omega  & 0 
\end{array}
\right),
\label{M2}
\end{eqnarray}

\begin{eqnarray}
{\bf M}_3\equiv\left(
\begin{array}{ccccc}
 0 & 0 & -A(0) & 0 & 0 \\
 0 & 0 & 0 & -A(0) & 0 \\
 0 & -B(0) & 0 & 0 & -A(0) \\
 -A(0) & 0 & 0 & 0 & 0 \\
 0 & -A(0) & -\omega  & 0 & -B(0) \\
 0 & \omega  & -B(0) & 0 & \omega  \\
 -\omega  & 0 & -\lambda  & -\omega  & 0 \\
 0 & -2 \omega  & A(0) & 0 & 0 \\
 0 & -\omega  & B(1) & 0 & -\omega  \\
 \omega  & 0 & -\lambda  & -\omega  & 0 \\
 A(1) & 0 & 0 & -\frac{1}{2} & 0 \\
 -2 \lambda  & C(1) & \omega  & -2 \lambda  & B(1) \\
 0 & 0 & C(1) & 0 & -2 \omega  \\
 -\frac{1}{2} & 0 & 0 & A(1) & 0 \\
 -2 \lambda  & B(1) & 2 \omega  & -2 \lambda  & C(1) \\
\end{array}
\right),
\label{M3}
\end{eqnarray}
and
\begin{equation}
\begin{aligned}
A(n)&\equiv-\lambda^2-\frac{n}{2}, \\
B(n)&\equiv-2\lambda^2-\frac{n}{2}, \\
C(n)&\equiv-3\lambda^2-\frac{n}{2}.
\end{aligned}
\end{equation}
The coefficients in stationary solution (\ref{eq:rhomatrixStaFeed}) read
\begin{equation}
\begin{aligned}
\Upsilon_1&\equiv 2 \left(32 \lambda ^{12}+82 \lambda ^{10}+\lambda ^8 \left(72 \omega ^2+65\right)+24 \lambda ^6 \left(4 \omega ^2+1\right)\right. \\
&+\left. \lambda ^4
   \left(48 \omega ^4+81 \omega ^2+4\right)+2 \lambda ^2 \omega ^2 \left(19 \omega ^2+9\right)+8 \left(\omega ^6+\omega ^4\right)\right),\\
\Upsilon_2&\equiv 2 \lambda  \left(2 \lambda ^2+1\right) \omega  \left(16 \lambda ^4+5 \lambda ^2+4 \omega ^2\right),\\
\Upsilon_3&\equiv -\omega  \left(32 \lambda ^8+90 \lambda ^6+\lambda ^4 \left(40 \omega ^2+57\right)+10 \lambda ^2 \left(3 \omega ^2+1\right)+8 \left(\omega
   ^4+\omega ^2\right)\right),\\
\Upsilon_4&\equiv -\left( 32 \lambda ^{10}+82 \lambda ^8+5 \lambda ^6 \left(8 \omega ^2+13\right)+6 \lambda ^4 \left(\omega ^2+4\right)+4 \lambda ^2 \left(2 \omega
   ^4+6 \omega ^2+1\right)+8 \left(\omega ^4+\omega ^2\right)\right) ,\\
\Upsilon_5&\equiv 4 \lambda  \omega ^2 \left(16 \lambda ^4+5 \lambda ^2+4 \omega ^2\right),\\
\Upsilon_6&\equiv \frac{1}{2}\left( 64 \lambda ^{12}+196 \lambda ^{10}+4 \lambda ^8 \left(36 \omega ^2+53\right)+\lambda ^6 \left(232 \omega ^2+113\right)\right.\\
 &+\left. 8 \lambda ^4
   \left(12 \omega ^4+21 \omega ^2+4\right)+\lambda ^2 \left(84 \omega ^4+60 \omega ^2+4\right)+8 \left(2 \omega ^6+3 \omega ^4+\omega
   ^2\right)\right),\\
\Upsilon_7&\equiv -\omega  \left(32 \lambda ^8+74 \lambda ^6+\lambda ^4 \left(40 \omega ^2+77\right)+\lambda ^2 \left(62 \omega ^2+32\right)+8 \omega ^4+12
   \omega ^2+4\right), 
\end{aligned}
\end{equation}
and
\begin{equation}
\begin{aligned}
T&\equiv 192 \lambda ^{12}+652 \lambda ^{10}+16 \lambda ^8 \left(27 \omega ^2+52\right)+\lambda ^6 \left(776 \omega ^2+551\right)+\lambda ^4
   \left(288 \omega ^4+716 \omega ^2+209\right)\\
   &+\lambda ^2 \left(268 \omega ^4+234 \omega ^2+44\right)+48 \omega ^6+64 \omega ^4+20 \omega
   ^2+4.
\end{aligned}
\end{equation}

%%%%%%%%%%%%%%%%%%%%%%%%%%%%%%%%%%%%%%%%%%%%%%%%%%

\end{document}